\documentstyle[12pt,aasms4]{article}

\begin{document}

\title{Optical/Near--Infrared Observations of GRO~J1744-28}

\author{D. M. Cole, D. E. Vanden~Berk, S. A. Severson, 
    M. C. Miller\altaffilmark{1}, J. M. Quashnock\altaffilmark{1},
    R. C. Nichol, D. Q. Lamb}
\affil{Department of Astronomy and Astrophysics \\
        University of Chicago, Chicago, IL 60637}

\author{K. Hurley}
\affil{Space Sciences Laboratory, \\
	University of California, Berkeley, CA 94720-7450}

\author{P. Blanco}
\affil{Center for Astrophysics and Space Studies \\
	University of California, San Diego \\
	La Jolla, CA 92093}

\author{C. Lidman}
\affil{European Southern Observatory \\
	Santiago, Chile}

\and \author{K. Glazebrook}
\affil{Anglo-Australian Observatory \\
        PO Box 296,
	Epping, NSW 2121, Australia}

\altaffiltext{1}{{\it Compton} Gamma-Ray Observatory Fellow}

\begin{abstract}

We present results from a series of optical (g and r-band) and
near--infrared (K$\arcmin$-band) observations of the region of the
sky including the entire XTE and ROSAT error circles 
for the ``Bursting Pulsar'' GRO~J1744-28.
These data were taken with
the Astrophysical Research Consortium's 3.5-m telescope at
Apache Point Observatory
and with the 2.2-m telescope at the European Southern
Observatory.
We see no new object, nor
any significant brightening of any known object,
in these error circles, with the exception of an object
detected in our 8 February 1996 image.
This object has already been proposed as a near--infrared
counterpart to GRO~J1744-28.  While it is seen in only two
of our ten 8 February frames,
there is no evidence that this is an instrumental artifact,
suggesting the possibility of
near--infrared flares from GRO~J1744-28, similar to those
that have been reported from the Rapid Burster.
The distance to the ``Bursting Pulsar'' must be more than 2 kpc,
and we suggest that it is more than 7 kpc.

\end{abstract}


\section{Introduction}

On December 2, 1995, a new source of X-ray bursts
lying in the direction of the galactic center was discovered
using the Burst and Transient Source Experiment (BATSE)
aboard the Compton Gamma-Ray Observatory
(\cite{fis95}; \cite{kou96b}).
This source, GRO~J1744-28, was an unusually frequent burster,
with a maximum rate of 18 per hour on December 2
(\cite{kou96b}) and an estimated rate of 40 per day
at its brightest (\cite{gab96b}).
Subsequent BATSE observations showed that GRO~J1744-28 is
also a persistent emitter of X-rays (\cite{pac96}),
and found pulsations at 2.1~Hz both in
the persistent emission (\cite{fin96a}) and in the
bursts (\cite{kou96a}),
making GRO~J1744-28 the
only known source of both periodic X-ray pulsations and
frequent X-ray bursts and resulting
in it being called the ``Bursting Pulsar''.
Timing of the 2.1~Hz oscillations
in the persistent X-ray emission revealed regular phase
shifts with a period of 11.83 days, which have been 
interpreted as orbital motion in a binary (\cite{fwvp96}).
The low inferred mass function 
($f_x(M)=1.36\times 10^{-4}\,M_\odot$, \cite{fin96b})
and eccentricity ($e<1.1\times 10^{-3}$, \cite{fin96b})
suggest that the neutron star in GRO~J1744-28 is being fed
by Roche lobe overflow from a low-mass red giant
(\cite{dau96}; \cite{lmt96a}; \cite{sd96}; \cite{bb96}; \cite{jr96}).

The optical brightness of many low-mass X-ray binaries (LMXBs)
is thought to be dominated by
reprocessing, in the accretion disk, of X-rays emitted from
the surface of the neutron star.  The counterpart to a
transient LMXB is thus often identified by the detection of
simultaneous brightening in other wavelengths (van~Paradijs \&
McClintock 1994, 1995).  Application of the
empirical formulae of van~Paradijs \& McClintock (1994, 1995)
indicates that if GRO~J1744-28 is at the galactic center,
the apparent unreddened visual magnitude
due to reprocessing is approximately 12.5.
An extension of the van~Paradijs \& McClintock fits to the
infrared suggests that at
8 kpc the unreddened K$\arcmin$ magnitude
would be approximately
14 (\cite{lmt96a}), making GRO~J1744-28 a good candidate
for optical or infrared identification.  Unfortunately,
GRO~J1744-28 was discovered when nearing closest approach
to the Sun, making it temporarily impossible for any
ground or space-based optical or infrared observatories to
view it.  Other transient LMXBs have active lifetimes of
only a few months; the expected early demise of GRO~J1744-28,
coupled with its unique nature, led us to undertake an
extensive series of observations in the optical and
near-infrared during January and February, 1996, in hopes of
finding a counterpart while it was possible.

Here we report our optical and infrared observations of 
GRO J1744-28.  In \S\ 2 we list the instruments and
modes of operation we used during the search, and present
our results for the infrared and optical variability
in the entire Rossi X-ray Timing Explorer (RXTE) and ROSAT
error boxes.
In \S\ 3 we discuss these results in light of models for the binary
system in GRO~J1744-28 and models for X-ray reprocessing,
and place lower limits on the distance to GRO~J1744-28.
We also consider the near--infrared counterpart proposed
by Augusteijn et al. (1996a);
we show there is no reason to believe this object is an artifact,
and that if it is associated with GRO~J1744-28,
the ``Bursting Pulsar'' exhibits flaring
behavior similar to that of the Rapid Burster.
Finally, in \S\ 4 we give our conclusions.

\section{Observations}

\subsection{Observational Procedure}

Optical observations were made with the ARC 3.5-m telescope
at Apache Point Observatory in New Mexico,
using the Double Imaging Spectrograph (DIS).
Via a dichroic (split at 5350 \AA), this camera can
observe simultaneously in both
blue and red, through Gunn g and r filters.
The blue chip is a 512x512
SITe CCD and the red chip an 800x800 TI array, with pixel scales of 1\farcs1
(blue) and 0\farcs61 (red); the usable areas of the
resulting images are approximately 6\farcm5~x~4\farcm5 (blue) and
5$\arcmin$~x~4\farcm5 (red).

Near--infrared observations were made with both the ARC
and the European Southern Observatory's 2.2-m telescope.
On the ARC, we used the Near Infrared Grism
Spectrometer and Imager II (GRIM II), a 256~x~256 NICMOS array,
with a
Mauna Kea K$\arcmin$ broadband filter (bandpass 1.95--2.30 $\mu$m).
At f/5, the pixel scale is 0\farcs47 and results in
a 2$\arcmin$~x~2$\arcmin$ field of view.  The skies were bright enough
in the infrared throughout our search that we were forced to use
neutral density (ND) filters.  For most of our observations
we used a 3\% transmission filter, but in March we switched to a 25\% one.

On the ESO 2.2-m, we used the IRAC2 camera, which
is also a 256~x~256 NICMOS array, again with a Mauna Kea K$\arcmin$
broadband filter.
Images were taken using pixel scales of both 0\farcs278 and
0\farcs507, giving fields of view of about
1$\arcmin$ and 2$\arcmin$ on a side, respectively.  

When we began observing GRO~J1744-28 on 12 January 1996, it did not
rise until after astronomical twilight had begun, and was only
25$\arcdeg$ from the Sun.  The minimum elevation of the ARC 3.5-m
is 6$\arcdeg$; by the time GRO~J1744-28 was high enough to be
observed, the Sun was less than $10\arcdeg$ below the horizon.
This meant that the sky was already too bright to observe anywhere
but in the near--infrared, and there were only 45 minutes until sunrise.
Each day thereafter, however, the time between sunrise and when
GRO~J1744-28 became visible grew by nearly 4
minutes, and the sky at acquisition became darker;
thus, by the beginning of February we were able to take relatively
deep optical images in addition to the infrared.

\subsection{Observational Log}

We summarize our
observations in Table 1.  Our first observations were taken on
the morning of 12 January 1996, when the best
position we had for GRO~J1744-28
was a parallelogram synthesized from Ulysses/GRB and GRO/BATSE
observations 
measuring roughly $24\arcmin$ by $7\arcmin$
(\cite{hur96}).  The central $10\arcmin$ by $6\arcmin$ portion
of this parallelogram
was covered with a 14-tile mosaic in K$\arcmin$;
unfortunately, the
eventual ROSAT position proved to be
just off the edge of the easternmost tile.

We next observed on the morning of 21 January, by which date
GRO~J1744-28 was visible for a few minutes
before the beginning of astronomical twilight.  Its position had also
been refined by observations made with the 
Rossi X-ray Timing Explorer (RXTE)
to a circle approximately 2$\arcmin$ in radius (Swank 1996).
We were only able to cover about 25\% of this
error circle, with a mosaic of 2 tiles.
Each tile is composed of
two 5-second integrations through a 3\% ND filter, giving the mosaic
a 3-$\sigma$ detection limit of magnitude 14.4 (Fig. 1a).

On 24 January we observed in the optical for the first time,
generating g and r-band stacks
of 28 and 37 seconds, respectively.
The r-band image covers the entire RXTE error circle,
and the g-band image the northern three-quarters;
both cover the ROSAT position discussed below.
The images have 3-$\sigma$
detection limits of 15.5 (g) and 17.0 (r).

Our next observations were on 30 January.  By that date,
the RXTE position for GRO~J1744-28
had shrunk to a circle approximately 1$\arcmin$ in radius
(\cite{str96}), meaning that
the entire error circle lay within a single GRIM II field of view.
Thus, rather than having to map out the error circle, we were able to
simply stare at the region and try to catch the source in a burst.
Ten-second K$\arcmin$ images
(still through a 3\% ND filter) were taken every 15 seconds
for over 30 minutes, from 12:42 to 13:13 UT; the telescope was dithered
by 20$\arcsec$ every five frames.
An X-ray burst was recorded by Ulysses/GRB at 13:08:38,
but did not coincide with any one image.  Our
previous frame ended at 13:06:37, the next began
at 13:09:01, and neither contained any flare.

The seeing was extremely variable during the 30 January observations.
In fact, we could use only 53 of the 100 frames
to construct a stacked image.  However, these 53 frames give us
an image equivalent to a 530-second integration
with a 3-$\sigma$ detection limit of magnitude 15.2
(Fig. 1b).

To complement the K$\arcmin$ data,
we took more optical images of the field on 5 February.
By then,
the separation of GRO~J1744-28 and the Sun was sufficient that
we were able to take 15-second exposures for a period
of about 90 minutes, dithering by about 20$\arcsec$ between
three positions.  We constructed
stacked images in g and r of
665 seconds each (Fig. 2), with 3-$\sigma$ detection limits
of magnitudes 20.5 and 19.7, respectively.

We made our first southern hemisphere observations on 8 February.
Using the ESO 2.2-m, we took ten 60-second K$\arcmin$ frames, each
dithered by 20\arcsec.  The stacked image has
a 3-$\sigma$ detection limit of 16.75 (Fig. 3).
Due to the southern location of ESO, the airmass was much less
than in the observations from Apache Point.  The seeing in
our ESO data is therefore much better ($\sim$ 0\farcs6),
and the limiting magnitude is significantly improved.

Our work in January and February was predicated on the idea that
like many transient LMXBs,
the ``Bursting Pulsar'' would quickly fade from view, and therefore
an immediate counterpart search was essential.
By the end of February, though, GRO~J1744-28
was well separated from the Sun and
still strongly emitting and bursting in X-rays,
and the field containing GRO~J1744-28 was being routinely
observed in many wavelengths.
Still, it seemed likely that after
three months, GRO~J1744-28 would soon decline; therefore,
on 3 March we observed again with the ARC 3.5-m to
create one last K$\arcmin$ image.
The sky by then was dark enough for us to use a 25\% ND
filter instead of the 3\% filter, speeding things up considerably.
Our stack of 54 5-second frames produced an image with a
3-$\sigma$ detection limit of magnitude 16.3
(Fig. 4a).

In early March, GRO~J1744-28 became
sufficiently separated from the Sun for ASCA and ROSAT to observe it.
The ASCA observations returned a position, good to 1$\arcmin$, which
partially overlaps the RXTE error circle, but
is shifted by more than 1\farcm5 to the northwest (\cite{dot96}).
Shortly thereafter, ROSAT determined a position good to 8$\arcsec$,
consistent with the ASCA position and just outside
the RXTE error circle
(\cite{kou96c}).\footnote{At the 1996 HEAD meeting,
J. Greiner of the ROSAT team (MPE Garching) indicated that the
inclusion of systematic errors boosts the error radius to 10$\arcsec$.}
Most of our observations were based on the RXTE positions,
and while they repeatedly cover its entire 1$\arcmin$ radius error circle,
the shift to the ROSAT position is enough to move GRO~J1744-28
off the edge of many of our K$\arcmin$ frames due to
the small fields of view of both the GRIM II and IRAC2 cameras.
Still, the ROSAT error circle is covered
by at least some K$\arcmin$ frames
on every night we observed, as well as by all of our
optical observations.

Although the ``Bursting Pulsar'' was active longer than we expected,
the persistent
emission from GRO~J1744-28 declined continuously from January to
March.  In April, GRO~J1744-28 was predicted to be undetectable by
early May (\cite{gab96a});\footnote{In fact, RXTE was still detecting
GRO~J1744-28, as of August 30, 1996.}
indeed,
it ceased being observable by GRO/BATSE on 3 May (\cite{kou96d}).
We therefore arranged to observe again
with the ESO 2.2-m in order to obtain a deep baseline infrared image.
We took 36 30-second K$\arcmin$ frames on 2 May, giving us an image
with a 3-$\sigma$ detection limit of 17.1, and an expanded plate
scale of 0\farcs278/pixel (Fig. 4b).

\subsection{Results}

To investigate the existence of any new or brightened objects in
the optical in the error circles for GRO~J1744-28,
we visually compared our images with Palomar Sky Survey
prints and ESO copies of UK Schmidt plates.  We also overlaid
our images with the digitized COSMOS/NRL source list,
concentrating on the area within 2\farcm5 of the RXTE position.
In the infrared, we began by blink-comparing our observations
with a 1992 NOAO set of J, H, and K images
kindly provided by Mike Merrill
(Merrill \& Gatley, private communication).  The 3-$\sigma$
detection limit of this K image is about 14,
at least a magnitude brighter
than the images we took after 30 January.  Therefore, we
also blink-compared
each of our infrared images with
our deepest image (taken on 2 May),
concentrating on the ROSAT error circle.
These comparisons are summarized in Table 2.

We find no new objects nor any objects which have
brightened by more than 0.5 magnitude
in any optical observation.
In the infrared, no objects differ
from either the 1992 NOAO K-band image or our 2 May image,
with one exception: our 8 February image contains an
object, proposed as the near--infrared counterpart to GRO~J1744-28 by
Augusteijn et al. (1996a)
through their comparison with their own March 28 K$\arcmin$ data
(but see \cite{aug96b} and \S\ 3).
Using STScI Digitized Sky Survey scans and the IRAF/STSDAS GASP package,
we find the position of this object to be
$17^{\rm{h}}44^{\rm{m}}33\fs05 \pm 0\fs02$,
-$28\arcdeg44\arcmin18\farcs6 \pm 0\farcs1$,
placing it just on the edge of the ROSAT error circle.
We also find the K$\arcmin$ magnitude of
the object to be $15.7\pm0.3$, in agreement
with that originally reported (\cite{aug96a}).

\section{Discussion}

\subsection{Distance constraints}

By combining our observations with models for GRO~J1744-28,
we can place limits on the distance to
GRO~J1744-28 and predict the K$\arcmin$ magnitude at which it must
be seen (see \cite{lmt96a} for details).
The companion in GRO~J1744-28 is believed to be a low-mass giant
that is transferring material onto the neutron star via Roche
lobe overflow
(\cite{dau96}; \cite{lmt96a}; \cite{sd96}; \cite{bb96}; \cite{jr96}).
The intrinsic luminosity and effective temperature of
the companion are then expected to be $20-30\,L_\odot$ and
$T\approx 4300$ K (se, e.g., \cite{lmt96a}).  The 
brightest near--infrared source in the ROSAT error box has
K$\arcmin$=11 and
there are no optical sources brighter than r=19.7.  Given
the expected luminosity and temperature of the companion, the 
lower distance limits (assuming $A_V\approx 3$ mag kpc$^{-1}$ and
$A_{K^\prime}\approx A_V/9$ (\cite{mat90}; \cite{dra93}))
are 1.5 kpc from the infrared limit and 2 kpc from the optical limit.

Another, somewhat more uncertain, limit may be derived
by using the van~Paradijs \& McClintock (1994, 1995) relation
between the X-ray luminosity and optical luminosity of
low-mass X-ray binaries.  For GRO~J1744-28 their relation
gives an absolute visual magnitude of $M_V\approx -2$.  Again
assuming a reddening of $A_V\approx 3$ mag kpc$^{-1}$ and
$A_{K^\prime}\approx A_V/9$, our limit
that no source had brightened at r=19.5 or brighter means
that the distance to GRO~J1744-28 must be greater than 3~kpc.  An
extension of the van~Paradijs \& McClintock relation
to the infrared (\cite{lmt96a}), combined with our limit
of $m_{K^\prime}=14$ for infrared brightening, gives a 
lower limit to the distance of 5 kpc.  An entirely 
independent distance limit consistent with our lower
limit was derived by Daumerie
et al. (1996), who used the standard theory of disk
accretion onto magnetized stars (\cite{gl79}) to
estimate the peak luminosity of GRO~J1744-28.  Combined
with the peak observed flux, this model gives a distance
greater than $\sim 7$~kpc.  A distance of $\sim$8~kpc is
also supported by the angular proximity of GRO~J1744-28
to the galactic center (only 20$\arcmin$ away) and by the
high neutral column density inferred from ASCA observations
(\cite{dot96}).  There is thus strong evidence that GRO~J1744-28
is near the galactic center.

\subsection{A Counterpart?}

The object proposed by Augusteijn et al. (1996a)
as the near--infrared counterpart to GRO~J1744-28
is seen only in our 8 February image;
we refer to it as the infrared candidate (IRC).
The 8 February image is a stack of ten 60-second K$\arcmin$ frames,
each frame dithered by $\sim 10\arcsec$ to facilitate flatfielding.
To further investigate the IRC, we examined
the frames individually (Table 3 and Fig. 5).
Of the ten frames, three were dithered
such that the location of the IRC is off the edge of the
array, and of the remaining seven frames,
the IRC is seen in only two; this
raises the possibility that the IRC is actually an artifact
(\cite{aug96b}).

The IRAC2 NICMOS chip has defects, seen in the flatfield,
which cause stars to fluctuate artificially.  The IRC does not,
however, coincide with any such defect in any frame.
Moreover, a chip defect must move relative to the sky as the
telescope is dithered---such an artifact can be seen in frames
8 and 9 of Figure 5---yet in these two frames the
IRC is seen at the same position relative to nearby stars, although
the frames were dithered by more than 20$\arcsec$ (40 pixels).

The seven frames in which the IRC could have been
seen have 3-$\sigma$ detection limits of magnitude 15.5--15.6.
With magnitude 14.6$\pm$0.4 in frame 4
and 15.0$\pm$0.4 in frame 8, the IRC is well above the
detection limit; it is brighter, in fact, than two neighboring stars.
In Figure 6 we compare the radial profile of the IRC (summed
over both frames) with a point spread function constructed from the
images of 10 stars of medium brightness.  The two are identical, within
statistical noise; the IRC is not a single-pixel event.

If we examine the frames in the order in which they were taken, we
see that the two frames in which the IRC is seen bracket the
three in which it could not possibly have been seen (see Fig. 5).
Thus, the two frames are neither isolated single frames nor
contiguous.  The detection of the IRC in those two frames is
consistent either with an event covering
5 frames, and lasting at most 6 minutes, or with two or more events
of average duration less than 3 minutes.

No X-ray bursts were recorded by Ulysses/GRB during the entire
period of integration (09:26$-$09:42 February 8 UT),
so if the IRC is an infrared burst or group of bursts from the
companion to GRO~J1744-28, the bursts are not correlated with
X-ray bursts.  We note, however, that
infrared flares a few minutes apart, uncorrelated 
with X-ray bursts, have been reported from the Rapid Burster 
(Apparao et al. 1979; Kulkarni et al. 1979; Jones et al.
1980).  The flares from the Rapid Burster are separated
by between 50 and 150 seconds, with total energies
(assuming isotropic emission and a distance of 10~kpc)
of between $\sim 6\times 10^{37}$ ergs and
$\sim 3\times 10^{38}$ ergs.

We conclude that there is no convincing instrumental reason
to doubt the reality of the images seen on these two 8 February
frames, and no obvious astrophysical reason that such images could
not be related to GRO~J1744-28.
Figure 7 shows the light
curve for the IRC over the entire period of our observations
(21 January -- 2 May 1996) and (see inset)
during the 8 February 1996 exposure.
Further observations are necessary; in particular, if
the IRC is the counterpart, a persistent source at this
location is expected with $m_{K^\prime}<20$ (\cite{lmt96a}).

If the IRC is an infrared burst or group of bursts
from GRO J1744-28, the minimum total energy in the bursts
may be estimated by comparing this with the 2 May image, in which the
companion must be fainter than 
$m_{K^\prime}\approx 17.1$.  Assuming isotropic emission
and using a companion luminosity of 20 $L_\odot$, the
average luminosity of the flare in the two frames in which
it was detected was at least $L\approx 250\,L_\odot$.  If the
flares only occurred in two one-minute frames, this 
implies a total energy of $\sim 10^{38}$ ergs, similar to
that seen from the Rapid Burster.  This is
a minimum, because in principle the source could have
been arbitrarily bright in the three intervening frames,
where the candidate was out of the field of view.

In addition, if the IRC is the counterpart, we can place two more
lower limits on to the distance to GRO J1744-28.
In our 2 May image, nothing is seen at the location of the
IRC to $m_{K^\prime}=17.1$; combined with the luminosity estimates
for the companion, this implies a distance of more than 7~kpc.
Similarly, if the IRC is the counterpart,
the fact that no sources were seen at the location of
the IRC to $m_{K^\prime}=15.5$
in five 8 February frames
implies a distance of $\sim$7~kpc.  The main
uncertainty in both estimates is the amount of reddening
to the galactic center, which could be between 
$A_{K\arcmin}=2$ and $A_{K\arcmin}=4$ (\cite{mat90}).

\section{Conclusions}

We observed the region of the sky including
the entire RXTE and ROSAT error circles for GRO~J1477-28 multiple
times in both the infrared (K$\arcmin$-band)
and in the optical (g and r-bands).  These observations allow
allow us to put strict lower limits on the distance to GRO~J1744-28
of 1.5 kpc (infrared) and 2 kpc (optical),
or, depending on the X-ray reprocessing model chosen,
5 kpc (infrared) and 3 kpc (optical).
Our 8 February observations show a possible near--infrared counterpart
to the ``Bursting Pulsar'' at
$17^{\rm{h}}44^{\rm{m}}33\fs05 \pm 0\fs02$,
~-$28\arcdeg44\arcmin18\farcs6 \pm 0\farcs1$;
if this is indeed the counterpart,
we can place a further limit on the distance to
GRO~J1744-28 of more than 7 kpc.
Additional observations are needed; in particular, a near--infrared image
of this field which reaches a limiting magnitude of K$\arcmin$=20.

\acknowledgments

We must acknowledge the extraordinary efforts made by the
observing specialists at APO, without whom this project would have
been impossible: E. Bergeron, K. Gloria, and D. Long.  DMC wishes to
thank his advisor, D. A. Harper, for allowing him to embark on this
diversion from his thesis.
The authors also wish to acknowledge the following sources of
support:
NASA Space Telescope Grant GO-06007.01-94 (DEVB),
Compton GRO fellowships NAGW 5-2687 (MCM) and NAGW 5-2660 (JMQ),
NASA grant NAG 5-2868 (DQL),
and JPL Contract 958056 for the Ulysses/GRB experiment (KH).

%
\clearpage
 
\begin{deluxetable}{rcccc}
\footnotesize
\tablecaption{Observations. \label{tbl-1}}
\tablewidth{0pt}
\tablehead{
\colhead{Date} & \colhead{UT}   & \colhead{Passband}   & \colhead{Exposure (sec)} & 
\colhead{Area Covered}
} 
\startdata
24 Jan &13:06$-$13:16  &g &28 &$17^{\rm{h}}44^{\rm{m}}22\fs0-51\fs8,~~$-$28\arcdeg41\arcmin00\arcsec-45\arcmin38\arcsec$ \nl
 & &r &37 &$17^{\rm{h}}44^{\rm{m}}19\fs7-54\fs2,~~$-$28\arcdeg41\arcmin14\arcsec-46\arcmin19\arcsec$ \nl
5 Feb &11:47$-$12:58  &g &665 &$17^{\rm{h}}44^{\rm{m}}17\fs2-43\fs0,~~$-$28\arcdeg42\arcmin38\arcsec-46\arcmin57\arcsec$ \nl
 & &r &665 &$17^{\rm{h}}44^{\rm{m}}21\fs2-40\fs1,~~$-$28\arcdeg42\arcmin52\arcsec-46\arcmin51\arcsec$ \nl
\nl
12 Jan &13:25$-$13:36  &K$\arcmin$ &10 &$17^{\rm{h}}43^{\rm{m}}45^{\rm{s}}-44^{\rm{m}}33^{\rm{s}},~~$-$28\arcdeg42\arcmin00\arcsec-48\arcmin24\arcsec$ \nl
21 Jan &13:36$-$13:39  &K$\arcmin$ &10 &$17^{\rm{h}}44^{\rm{m}}29\fs1-38\fs2,~~$-$28\arcdeg44\arcmin07\arcsec-47\arcmin42\arcsec$ \nl
30 Jan &12:42$-$13:13  &K$\arcmin$ &530 &$17^{\rm{h}}44^{\rm{m}}29\fs1-38\fs2,~~$-$28\arcdeg44\arcmin05\arcsec-46\arcmin04\arcsec$ \nl
8 Feb &09:26$-$09:38  &K$\arcmin$ &600 &$17^{\rm{h}}44^{\rm{m}}28\fs0-39\fs5,~~$-$28\arcdeg44\arcmin08\arcsec-46\arcmin38\arcsec$ \nl
3 Mar &12:37$-$12:47  &K$\arcmin$ &270 &$17^{\rm{h}}44^{\rm{m}}32\fs8-37\fs2,~~$-$28\arcdeg44\arcmin13\arcsec-45\arcmin37\arcsec$ \nl
2 May &07:43$-$08:16  &K$\arcmin$ &1080 &$17^{\rm{h}}44^{\rm{m}}29\fs3-37\fs4,~~$-$28\arcdeg43\arcmin33\arcsec-45\arcmin03\arcsec$ \nl
\enddata

\tablecomments{The 8 February and 2 May observations were done at ESO;
all the rest at APO.}

\end{deluxetable}

\clearpage
 
\begin{deluxetable}{cccccc}
\footnotesize
\tablecaption{Image Comparisons
\label{tbl-2}}
\tablewidth{0pt}
\tablehead{
\colhead{Date} & \colhead{Passband} &
\colhead{Region of Comparison} & \colhead{Area \sq\arcmin} &
\colhead{Magnitude Limit} & \colhead{Notes}
} 
\startdata
24 Jan &g  &$17^{\rm{h}}44^{\rm{m}}22\fs9-45\fs7,~~$-$28\arcdeg42\arcmin52\arcsec-45\arcmin38\arcsec$ &11.15  &15.5 $\pm$ 0.5 &a \nl
  &r  &$17^{\rm{h}}44^{\rm{m}}22\fs9-45\fs7,~~$-$28\arcdeg42\arcmin52\arcsec-46\arcmin19\arcsec$ &14.63  &17.0 $\pm$ 0.5 &b \nl
5 Feb &g  &$17^{\rm{h}}44^{\rm{m}}22\fs9-43\fs0,~~$-$28\arcdeg42\arcmin52\arcsec-46\arcmin57\arcsec$ &18.49  &20.5 $\pm$ 0.3 &a \nl
  &r  &$17^{\rm{h}}44^{\rm{m}}22\fs9-40\fs1,~~$-$28\arcdeg42\arcmin52\arcsec-46\arcmin51\arcsec$ &13.31  &19.7 $\pm$ 0.3 &b \nl
\nl
21 Jan &K$\arcmin$  &$17^{\rm{h}}44^{\rm{m}}28\fs8-37\fs4,~~$-$28\arcdeg44\arcmin10\arcsec-47\arcmin45\arcsec$ &6.75  & $\sim$14    &c \nl
  &K$\arcmin$  &$17^{\rm{h}}44^{\rm{m}}31\fs0-35\fs6,~~$-$28\arcdeg44\arcmin07\arcsec-45\arcmin03\arcsec$ &0.94  &14.4 $\pm$ 0.3   &d \nl
30 Jan &K$\arcmin$  &$17^{\rm{h}}44^{\rm{m}}29\fs1-38\fs0,~~$-$28\arcdeg44\arcmin10\arcsec-46\arcmin10\arcsec$ &3.90  & $\sim$14    &c \nl
  &K$\arcmin$  &$17^{\rm{h}}44^{\rm{m}}30\fs9-35\fs6,~~$-$28\arcdeg44\arcmin05\arcsec-45\arcmin11\arcsec$ &1.13  &15.2 $\pm$ 0.3   &d \nl
 8 Feb &K$\arcmin$  &$17^{\rm{h}}44^{\rm{m}}32\fs0-34\fs3,~~$-$28\arcdeg44\arcmin10\arcsec-44\arcmin45\arcsec$ &0.29  & $\sim$14    &c \nl
  &K$\arcmin$  &$17^{\rm{h}}44^{\rm{m}}30\fs8-35\fs6,~~$-$28\arcdeg44\arcmin08\arcsec-45\arcmin08\arcsec$ &1.05  &16.75 $\pm$ 0.3  &d \nl
 3 Mar &K$\arcmin$  &$17^{\rm{h}}44^{\rm{m}}32\fs8-34\fs3,~~$-$28\arcdeg44\arcmin10\arcsec-44\arcmin45\arcsec$ &0.19  & $\sim$14    &c \nl
  &K$\arcmin$  &$17^{\rm{h}}44^{\rm{m}}32\fs8-34\fs3,~~$-$28\arcdeg44\arcmin08\arcsec-44\arcmin45\arcsec$ &0.20  &16.3 $\pm$ 0.3   &d \nl
\enddata

\tablenotetext{a}{Compared with a POSS print and the COSMOS/NRL list. }
\tablenotetext{b}{Compared with an ESO copy of a UK Schmidt plate and the
COSMOS/NRL list.}
\tablenotetext{c}{Compared with a 1992 NOAO K image, limiting
magnitude approximately 14.}
\tablenotetext{d}{Compared with the 2 May observation,
limiting magnitude 17.1 $\pm$ 0.3.}

\tablecomments{
The second RXTE error circle is $17^{\rm{h}}44^{\rm{m}}34\fs3 \pm 2\fs8,
~$-$28\arcdeg45\arcmin22\arcsec \pm 47\arcsec$ \space (\cite{str96}),
while the ROSAT error circle is $17^{\rm{h}}44^{\rm{m}}33\fs1,
~$-$28\arcdeg44\arcmin29\arcsec$, both $\pm 10\arcsec$
\space (\cite{kou96c}).
The region of comparison for any
optical observation is really that portion of a 2\farcm5 radius
circle centered on the second RXTE position--the
extent of our COSMOS/NRL list--which overlaps the image.
}

\end{deluxetable}

\clearpage

\begin{deluxetable}{ccccc}
\footnotesize
\tablecaption{8 February 1996 Frames. \label{tbl-3}}
\tablehead{
\colhead{Frame} & & \colhead{Candidate} & \colhead{Limiting} &
\colhead{X \& Y Shifts} \cr
\colhead{Number} & \colhead{UT} & \colhead{Seen} & \colhead{Magnitude} &
\colhead{(pixels)}
} 
\startdata
  1  & 09:27$-$09:28  &  No      &  15.5 &  18.7, ~~0.4  \nl
  2  & 09:28$-$09:29  &  No      &  15.5 &  19.2, ~20.1  \nl
  3  & 09:30$-$09:31  &  No      &  15.6 &  ~0.4, ~20.0  \nl
  4  & 09:31$-$09:32  & Yes      &  15.5 &  ~0.0, ~~0.0  \nl
  5  & 09:33$-$09:34  & off top  &   -   &  ~2.0, -19.5  \nl
  6  & 09:35$-$09:36  & off top  &   -   &  23.3, -19.9  \nl
  7  & 09:37$-$09:38  & off top  &   -   &  42.9, -19.8  \nl
  8  & 09:38$-$09:39  & Yes      &  15.6 &  43.2, ~-0.8  \nl
  9  & 09:40$-$09:41  &  No      &  15.5 &  41.8, ~18.6  \nl
 10  & 09:41$-$09:42  &  No      &  15.5 &  21.8, ~~0.2  \nl
\enddata

\tablecomments{
The magnitudes of the IRC in frames 4 and 8 are
14.6 $\pm$ 0.4 and 15.0 $\pm$ 0.4, respectively.
}

\end{deluxetable}

\clearpage

\figcaption[]{
K$\arcmin$, 21 January 1996 (a) and 30 January 1996 (b), APO.
The ROSAT error circle is marked.
}

\figcaption[]{
g-band (a) and r-band (b), 5 February 1996, APO.
The ROSAT error circle is marked.
}

\figcaption[]{
K$\arcmin$, 8 February 1996, ESO.
The IRC is visible at the top of the ROSAT error circle
(see also Fig. 5).
}

\figcaption[]{
K$\arcmin$, 3 March 1996, APO (a) and 2 May 1996, ESO (b).
The ROSAT error circle is marked.
}

\figcaption[]{
Montage of the frames taken on 8 February.  The sequence
begins in the lower left corner, runs up the left column, then up
the right column.
The white arrows point to the IRC in frames 4 and 8.
The location of the IRC is off the edge of frames 5--7.
A chip artifact is visible in frames 8 and 9.
}

\figcaption[fig6.eps]{ Radial profile of the IRC.  The smooth curve shows the
point spread function, obtained from 10 stars of medium brightness.}

\figcaption[fig7.eps]{ The light curve for the IRC over the entire
period of our observations (21 January -- 2 May 1996)
and (see inset) during the 8 February 1996 exposure
(see Tables 2 and 3).
The open symbol is from Augusteijn et al. 1996a.
The dashed limit is for the 8 February frames in
which the IRC does not appear.}

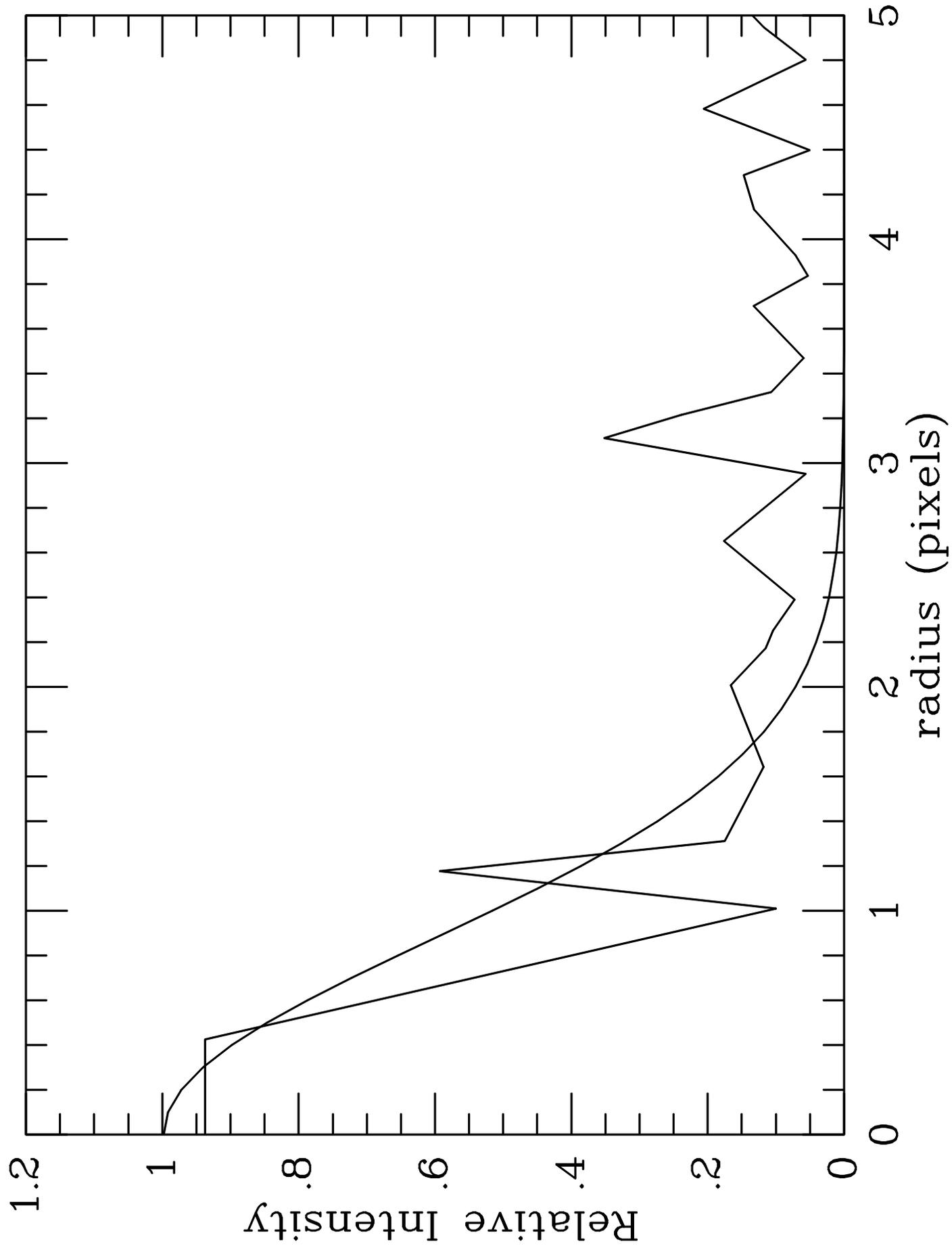
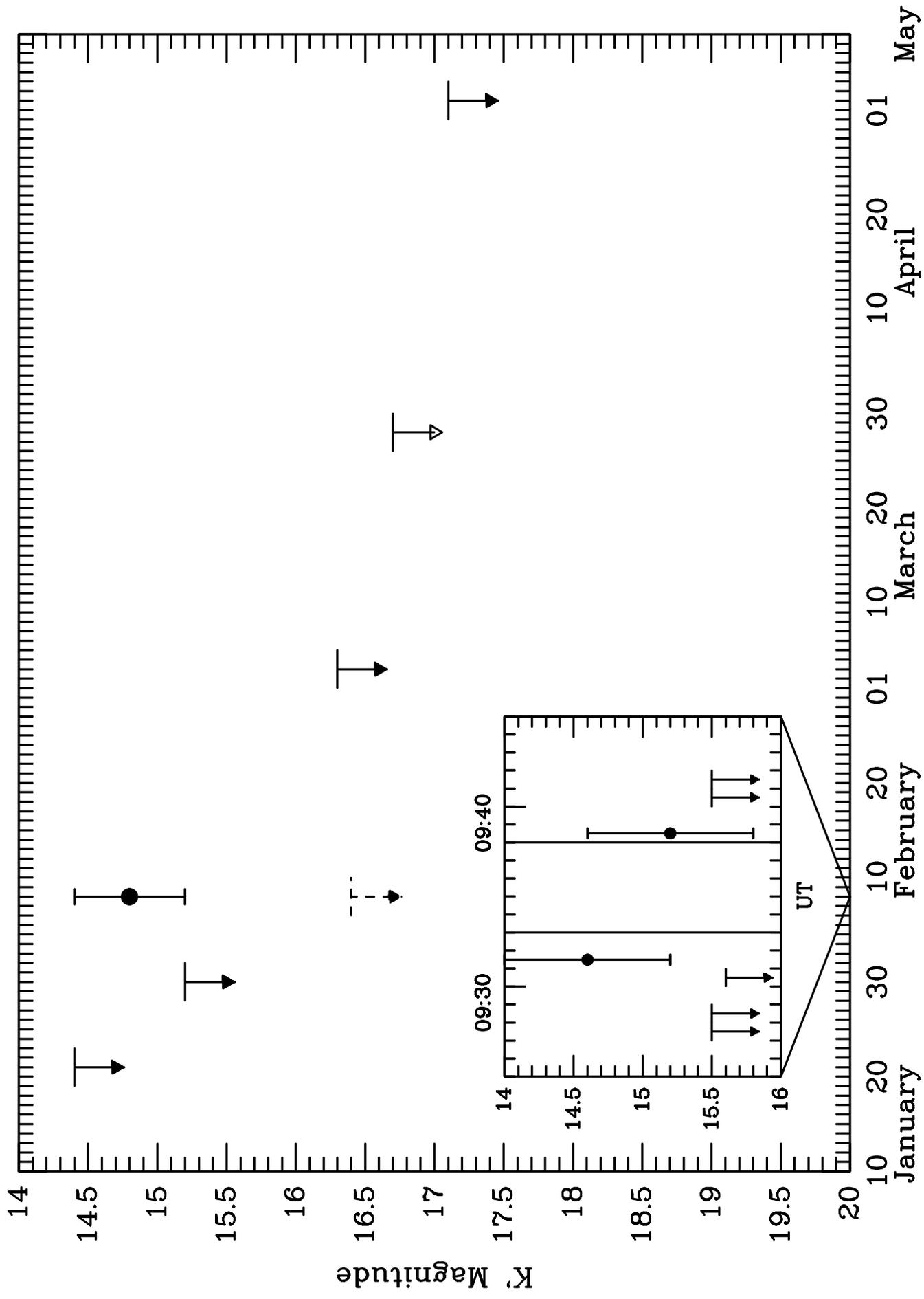
\end{document}